\documentclass[11pt,a4paper]{article}

\usepackage{jheppub}
\usepackage{amsfonts}
\usepackage{amsmath,graphicx,amsfonts,mathrsfs,amssymb}
\usepackage{amsmath,epsfig}
\usepackage{amssymb,amsfonts}
\usepackage{latexsym}
\usepackage{epsfig}
\newbox\pippobox
\def\be{\begin{equation}}
\def\ee{\end{equation}}
\def\bea{\begin{eqnarray}}
\def\eea{\end{eqnarray}}

\def\ee           {{\rm e}}

\def\tr           {\mathop{\rm Tr}}

\newcommand{\beq}{\begin{equation}}
\newcommand{\eeq}{\end{equation}}
\newcommand{\beqa}{\begin{eqnarray}}
\newcommand{\eeqa}{\end{eqnarray}}
\newcommand{\beqar}{\begin{eqnarray*}}
\newcommand{\eeqar}{\end{eqnarray*}}

\renewcommand{\eqref}[1]{(\ref{#1})}

\def\fR{{\mathfrak R}}

\catcode`\@=12

\title{Note on Integrability of Marginally Deformed ABJ(M) Theories}
\author[a,b]{Song He,}
\author[c]{Jun-Bao Wu}

\affiliation[a]{State Key Laboratory of Theoretical Physics,
Institute of Theoretical Physics, Chinese Academy of Science,
No.55 Zhongguancun East Road, Beijing 100190, P. R. China }\affiliation[b]{ Kavli Institute for
Theoretical Physics China, CAS,  No.55 Zhongguancun East Road, Beijing 100190, P. R. China}
\affiliation[c]{Institute of High Energy Physics, and Theoretical
Physics Center for Science Facilities, Chinese Academy of Sciences, 19B YuquanLu,
Beijing 100049, P.R. China}

\emailAdd{hesong@itp.ac.cn}\emailAdd{wujb@ihep.ac.cn}

\date{\today}

\abstract{We study the anomalous dimensions of operators in the
scalar sector of planar $\beta$-deformed ABJ(M) theories. We show that the
anomalous dimension matrix at two-loop order gives an integrable
Hamiltonian acting on an alternating $SU(4)$ spin chain with the spins
at odd lattice sides in the fundamental  representation and the
spins at even lattices in the anti-fundamental representation. We
get a set of $\beta$-deformed Bethe ansatz equations which give the
eigenvalues of Hamiltonian of this deformed spin chain system.
Based on our computations, we also extend our study to
non-supersymmetric three-parameter $\gamma$-deformation of ABJ(M)
theories and find that the corresponding Hamiltonian is the same as
the one in $\beta$-deformed case at two-loop level in the scalar
sector.}


\arxivnumber{1302.2208}
\begin{document}

\maketitle

\section{Introduction}

There were great achievements on integrable structure in both sides
of AdS/CFT correspondence in the last decade \cite{Beisert:2010jr}.
The best studied case is the integrability in
the correspondence between four-dimensional ${\cal N}=4$
super-Yang-Mills theory and the type IIB superstring theory on
$AdS_5\times S^5$ \cite{Mal97, Gubser:1998bc, Witten:1998qj}. In the
field theory side, it was found that, first at lower loop level, the
anomalous dimension matrix coincides with Hamiltonian of certain
integrable spin chain \cite{Minahan:2002ve, Beisert:2003tq, Beisert:2003jj}. The all-loop Bethe ansatz equations
\cite{Beisert:2003ys} were later
proposed and they can be obtained from a S-matrix for the spin
chain \cite{Beisert:2005tm}, though we do not know the corresponding Hamiltonian of the
spin chain at all-loop level. In the string theory side,  infinity
number of conserved charges on the worldsheet of Green-Schwarz
superstring moving in $AdS_5\times S^5$ were constructed \cite{Bena:2003wd}. The
integrable structure is a very powerful tool which
enables us to compute, as an example, the cusp anomalous dimension
for arbitrary value of the 't Hooft coupling in the planar limit
\cite{Beisert:2006ez, Freyhult:2010kc}.

Such integrable structure was also found for the recently proposed
duality \cite{ABJM, ABJ} between three-dimensional ${\cal N}=6$
Chern-Simons-matter theory and type IIA theory on $AdS_4\times CP^3$
\cite{MinahanZarembo}-\cite{Beccaria:2012vb} (for reviews see
\cite{Klose:2010ki, Lipstein:2011}). The dynamics in this case is
more complicated and richer than the previous one, partly due to the
fact that we now have less supersymmetries. As an example, there is
 still a to-be-determined function in the dispersion relation of
the magnon \cite{Gaiotto:2008cg, Grignani:2008is, Nishioka:2008gz}.
In the ${\cal N}=4$ super Yang-Mills case, this function is trivial
due to the fact that the theory is self-dual under the S-duality
transformation \cite{Berenstein:2009qd}.

With the success of integrability in mind, it should be with great
value to generalize the above studies to case with less
supersymmetries. The theory in which the anomalous dimensions of
gauge invariant operators are related to an integrable Hamiltonian
seems to be quite rare. If we perform generic marginal deformations
of the ${\cal N}=4$ super Yang-Mills theory \cite{Leigh:1995ep} or
ABJM theory, the obtained theory seems usually not to keep these
integrable structure in the above sense, even when some
supersymmetries and conformal symmetry are preserved. The $\beta$-
and $\gamma$-deformations of ${\cal N}=4$ super Yang-Mills theory
are quite special since they preserve this remarkable integrable
structure \cite{Roiban:2003dw}-\cite{Frolov:2005dj} and their
gravity dual can be obtained through a certain solution generating
technique \cite{Lunin:2005jy}.  Further studies of
this integrability can be found in
\cite{Frolov:2005ty}-\cite{Zoubos:2010kh}. These marginal
deformations are special also because they can be expressed elegantly
using a star product which produces a certain phase factor for each
interaction term in the Lagrangian. The solution generating
transformation in the gravity side can be constructed by
T-duality-shift-T-duality transformations in string theory \cite{Lunin:2005jy}. The
understand of the gauge-gravity correspondence in this case was
improved in  \cite{Imeroni:2008cr}. 
The $\beta$- and $\gamma$-
deformed  ABJM theories and their gravity duals were also studied in
\cite{Imeroni:2008cr}. Some classical string solutions in these
deformed backgrounds of type IIA string theory have been studied in
\cite{Schimpf:2009rk}-\cite{Ratti:2012kq}. The aim of this paper is
to explore the integrable structure in the field theory side of
these deformed $AdS_4/CFT_3$ correspondence.

We begin with the scalar sector of the $\beta$-deformed ABJ(M) theory which has ${\cal N}=2$ supersymmetries. 
We compute the anomalous dimensions of these operators at the
two-loop level in the planar limit and for all operators with length
larger than $2$. We express the result as a Hamiltonian of an
alternating $SU(4)$ spin chain. Comparing with the undeformed case
\cite{MinahanZarembo, Bak:2008cp, BakGangRey}, we find that, in the
Hamiltonian,  only the terms from the interaction terms with six
scalars are deformed. Though the interaction terms with two scalars
and two fermions are also deformed, their contributions to the
Hamiltonian coincide with the undeformed case. We also find that the
Hamiltonian in non-supersymmetric three-parameter $\gamma$-deformed
ABJ(M) theory is the same as the one of the $\beta$-deformed theory,
at two-loop level in the scalar sector. We expect the differences
will appear in other sectors and/or at higher loop order. As in
\cite{Beisert:2005if},
 we deform the  R-matrices constructed in \cite{MinahanZarembo, Bak:2008cp} by introducing suitable phase factors. We show that the obtained transfer matrices
 will produce essentially the Hamiltonian from the perturbative computations in the Chern-Simons-matter theories. This result shows that
 the Hamiltonian is integrable.  By diagonalizing the transfer matrices, we obtain the Bethe ansatz equations and the eigenvalues of the Hamiltonian.

The organization of the remaining parts of  this paper is as follows.
In section 2 we briefly review $\beta$-deformation of ABJM theory.
In section 3, we compute the two-loop corrections to the anomalous
dimensions of operators in the scalar sector in both ABJM and ABJ
theories. In section 4, we constructed the $R$ matrices after
deformation and show that the Hamiltonian obtained in section 2 is
integrable. Based on these results, we derive the eigenvalues of the
Hamiltonian of the deformed spin chain system in section 5. A brief
discussion on non-supersymmetric three-parameter
$\gamma$-deformation is put in section 6. Section 7 is devoted to
conclusions and discussions. We put some details of the computations
in section 4 in the Appendix.

{\bf Notes added in Nov. 2016:} Recently we realised that the $\gamma$-deformation discussed in section 6
of this paper is different from the one in \cite{Imeroni:2008cr}. The $\gamma$-deformation studied here is still well-defined and the discussion in section 7
is still valid. The integrability of the $\gamma$-deformed ABJM theory in \cite{Imeroni:2008cr} is recently studied in details in \cite{chenliuwu}.
We thank Hui-Huang Chen for discussions on this issue.

{\bf Notes added in Jan. 2021: } After the paper was published in JHEP, We corrected some errors especially the ones in the action of $\beta$-deformed ABJM theory in the e-print on arXiv. Main conclusion and results are unchanged.

\section{$\beta$-deformation of superconformal Chern-Simons-matter theory}
The ABJM theory \cite{ABJM} is three dimensional ${\cal N}=6$
supersymmetric Chern-Simons-matter theory. The gauge group of this
theory is $U(N)\times U(N)$, and the Chern-Simons levels of these two
subgroups are $k$ and $-k$, respectively. The matter fields are four
complex scalars $Y^I, I=1, 2, 3, 4$ and four fermions $\Psi_I, I=1,
2, 3, 4$ in the $(N, \bar{N})$ representation. The action of this
theory is\footnote{We follow the convention of \cite{Bak:2008cp}
closely.}

\bea S=\int d^3x (L_{CS}+L_{kin.}-V_F-V_B),\eea with \bea
L_{CS}=\frac{k}{4\pi}\epsilon^{\mu\nu\rho}\mbox{Tr}(A_\mu\partial_\mu
A_\rho+\frac{2i}{3}A_\mu A_\nu A_\rho)-
\frac{k}{4\pi}\epsilon^{\mu\nu\rho}\overline{\mbox{Tr}}(\bar{A}_\mu\partial_\mu
\bar{A}_\rho+\frac{2i}{3}\bar{A}_\mu \bar{A}_\nu \bar{A}_\rho),\eea
\bea L_{kin.}&=&\frac12\overline{\mbox{Tr}}(-(D_\mu Y)^\dagger D^\mu
Y^I+i\Psi^{\dagger I}\gamma_\mu D^\mu\Psi_I)+\frac12\mbox{Tr}(-D_\mu
Y^I (D^\mu Y)_I^\dagger \nonumber \\ &+&i\Psi_I\gamma_\mu D^\mu
\Psi^{\dagger I}), \eea
 \bea V_{\rm B} &=& - {1 \over 3} \left({2 \pi \over
k}\right)^2 \overline{\mbox{Tr}} \Big[ \, Y^\dagger_I Y^J
Y^\dagger_J Y^K Y^\dagger_K Y^I
+ Y^\dagger_I Y^I Y^\dagger_J Y^J Y^\dagger_K Y^K \nonumber \\
&& \hskip1.8cm + 4 Y^\dagger_I Y^J Y^\dagger_K Y^I Y^\dagger_J Y^K -
6 Y^\dagger_I Y^I Y^\dagger_J Y^K Y^\dagger_K Y^J \, \Big],
\label{Bpot} \eea \bea V_{\rm F} &=& {2 \pi i \over k}
\overline{\mbox{Tr}} \Big[ Y^\dagger_I Y^I \Psi^{\dagger J} \Psi_J -
2 Y^\dagger_I Y^J  \Psi^{\dagger I} \Psi_J  + \epsilon^{IJKL}
Y^\dagger_I \Psi_J Y^\dagger_K \Psi_L]
\nonumber \\
&-& {2 \pi i \over k} \mbox{Tr} [Y^I Y^\dagger_I \Psi_J
\Psi^{\dagger J} - 2 Y^I Y^\dagger_J \Psi_I \Psi^{\dagger J} +
\epsilon_{IJKL} Y^I \Psi^{\dagger J} Y^K \Psi^{\dagger L} \Big].
\label{Fbot} \eea 
The
covariant derivatives are: \bea D_\mu Y^I=\partial_\mu Y^I+i A_\mu
Y^I-iY^I\bar{A}_\mu, D_\mu Y^\dagger_I=\partial_\mu
Y^\dagger_I+i\bar{A}_\mu Y^\dagger_I-iY^\dagger_I \bar{A}_\mu, \eea
\bea D_\mu \Psi_I=\partial_\mu \Psi_I+i A_\mu
\Psi_I-i\Psi_I\bar{A}_\mu, D_\mu \Psi^{\dagger I}=\partial_\mu
\Psi^{\dagger I}+i\bar{A}_\mu \Psi^{\dagger I}-i\Psi^{\dagger I}
\bar{A}_\mu.\eea

In the following part, we will discuss the $\beta$ deformation of
the theory following the convention given by
\cite{Imeroni:2008cr}. The deformed theory will preserve three dimensional ${\cal N}=2$
supersymmetries. The $\beta$-deformation can be performed by replacing
all of the ordinary product $fg$ of two fields $f$ and $g$ in the
Lagrangian by the following star product:
\begin{equation}\label{starbeta}
     f * g = e^{i \pi \gamma (Q^f_1 Q^g_2 - Q^f_2 Q^g_1)} f g\,,
\end{equation}
where $Q^f_i, i=1, 2$ are two global $U(1)$ charges carrying by the
field $f$ and $\gamma$ is a real deformation parameter.\footnote{We denote the deformation parameter as $\gamma$ to stress that it is {\it real}, However
the supersymmetric one-parameter deformation is still called $\beta$-deformation and the non-supersymmetric three-parameter deformation to be introduced in section~6 will be called
$\gamma$-deformation. We hope this will not produce confusions for the readers.} For the
$\beta$-deformation of ABJM theory, we choose the charges for the
scalars as  in table~\ref{t:M2}. The fermionic super-partner
$\psi^{\dagger I}$ carries the same charge as $Y^I$, and the gauge
field is neutral under these symmetries.

It is easy to see that this star product is associative and for
several fields $F_1, ..., F_n$, we have \bea F_1 *...*F_n = e^{i \pi
\gamma \sum_{i<j} (Q_1^{F_i} Q_2^{F_j}- Q_2^{F_i} Q_1^{F_j})}
F_1...F_n. \eea This deformation just adds phase factors according
to the above equation to the interaction terms in the Lagrangian.
One can see from the above rule for the star product that the deformation does not change the
kinetic terms of the Lagrangian.
only deform the superpotential $W$  in the following manner given in  

\begin{table}
  \centering
  \caption{{$U(1)^2$ charges of the scalars of the ABJM
theory used for $\beta$-deformation.}}\label{t:M2}
\begin{tabular}{|c|c|c|c|c|}
\hline
& $Y^1$ & $Y^2$ & $Y^3$ & $Y^4$ \\
\hline
$U(1)_1$ & $+\frac{1}{2}$ & $-\frac{1}{2}$ & $0$ & $0$ \\
\hline
$U(1)_2$ & $0$ & $0$ & $\frac{1}{2}$ & $-\frac{1}{2}$ \\
\hline
\end{tabular}
\end{table}

For later use, we now give the interaction terms after the
deformation. One can see that only $V_B$ and $V_F$ will be deformed.
By a bit calculations, we found that the third term in scalar
potential, eq.~(\ref{Bpot}), will be deformed by multiplying the
phase factor $\exp{(-2 i \pi \gamma(Q^I\times Q^J+Q^I \times Q^K +
Q^K\times Q^I))}$. Here $I, J, K$ take value of the integral number
$1,2,3,4$, and we define:
\begin{equation}  Q^I\times Q^J\equiv
Q^I_1Q^J_2-Q^J_1Q^I_2.\end{equation} Where $Q^I$ are the $U(1)$
charges of the fields.

\begin{table}
  \centering
  \caption{The non-vanishing phase factors for the third term of $V_B$}\label{charge}
\begin{tabular}{|c|c|c|c|c|}
\hline
 $I$ & $J$ & $K$ & phase factor \\
\hline
$1$ & $2$ & $3$ & $i \pi \gamma$\\
\hline
$1$ & $2$ & $4$ & $-i \pi \gamma$\\
\hline
$1$ & $3$ & $2$ & $-i \pi \gamma$\\
\hline
$1$ & $3$ & $4$ & $-i \pi \gamma$\\
\hline
$1$ & $4$ & $2$ & $i \pi \gamma$\\
\hline
$1$ & $4$ & $3$ & $i \pi \gamma$\\
\hline
\end{tabular}
\end{table}

So the third term now becomes\footnote{In previous version of this paper, we included an extra term in the RHS of this equation incorrectly. After Nan Bai pointed to us, we realized that it is part of the first term. So we deleted it to avoid double counting. We thanks Nan Bai for pointing this to this. }: \bea && V_{\text{B,
3rd}}^{\text{deformed}}\nonumber\\&=& - {1 \over 3} \left({2 \pi
\over k}\right)^2 \left[4\sum_{\text{two of I, J, K are the
same}}\overline{\mbox{Tr}}(Y_I^\dagger Y^JY_k^\dagger Y^IY_J^\dagger Y^K) \right.
\nonumber \\&+&4\sum_{(IJK)=(123),(143),(142)}e^{i\pi \gamma}
\overline{\mbox{Tr}}(Y_I^\dagger Y^JY_k^\dagger Y^IY_J^\dagger Y^K)+\text{cyclic
permutations}\nonumber\\&+&\left.
4\sum_{(IJK)=(132),(134),(124)}e^{-i\pi
\gamma}\overline{\mbox{Tr}}(Y_I^\dagger Y^JY_k^\dagger Y^IY_J^+\dagger Y^K)+\text{cyclic
permutations}\right].\eea The non-vanishing phase factors  in the third term of the potential $V_B$ are listed in
table~\ref{charge}. The other three terms in eq.~(\ref{Bpot}) are untouched by the
$\beta$-deformation.


\begin{table}
  \centering
  \caption{The non-vanishing phase factors for the second term of $V_F$}\label{chargef1}
\begin{tabular}{|c|c|c|c|c|}
\hline
 $I$ & $J$  & phase factor \\
\hline
$1$ & $3$  & $-{1\over 2}i \pi \gamma$\\
\hline
$1$ & $4$  & ${1\over 2}i \pi \gamma$\\
\hline
$2$ & $3$  & ${1\over 2}i \pi \gamma$\\
\hline
$2$ & $4$  & $-{1\over 2}i \pi \gamma$\\
\hline
$3$ & $1$  & ${1\over 2}i \pi \gamma$\\
\hline
$3$ & $2$ & $-{1\over 2}i \pi \gamma$\\
\hline
$4$ & $1$ & $-{1\over 2}i \pi \gamma$\\
\hline
$4$ & $2$  & ${1\over 2}i \pi \gamma$\\
\hline
\end{tabular}
\end{table}

\begin{table}
  \centering
  \caption{The non-vanishing phase factors for the third term of $V_F$}\label{chargef2}
\begin{tabular}{|c|c|c|c|c|}
\hline
 $I$ & $J$  & $K$ & $L$  & phase factor \\
\hline
$1$ & $3$  &$2$ & $4$  & ${1\over 2}i \pi \gamma$\\
\hline
$1$ & $4$  &$2$ & $3$  & $-{1\over 2}i \pi \gamma$\\
\hline
$2$ & $3$  &$1$ & $4$  & $-{1\over 2}i \pi \gamma$\\
\hline
$2$ & $4$  & $1$ & $3$  &${1\over 2}i \pi \gamma$\\
\hline
$3$ & $1$  &$4$ & $2$  & $-{1\over 2}i \pi \gamma$\\
\hline
$3$ & $2$ & $4$ & $1$  &${1\over 2}i \pi \gamma$\\
\hline
$4$ & $1$ &$3$ & $2$  & ${1\over 2}i \pi \gamma$\\
\hline
$4$ & $2$  &$3$ & $1$  & $-{1\over 2}i \pi \gamma$\\
\hline
\end{tabular}
\end{table}

Now we turn to consider the interaction terms between the fermions
and scalars in eq.~(\ref{Fbot}). After the deformation, these terms
become\bea V_F^{\text{deformed}}&=&{2\pi i\over k}
\overline{\mbox{Tr}}
\Big[Y_I^\dagger Y^I\Psi^{\dagger J}\Psi_{J}-2\sum_{\text{$I=J$, or $I,J \in\{1,
2\}$, or $I,J \in\{3,
4\}$}}Y_I^\dagger Y^J\Psi^{\dagger I}\Psi_{J}\nonumber\\&&-2\sum_{(I, J)=(1, 4),
(2, 3), (3, 1), (4, 2)}e^{{i\over2}\pi
\gamma}Y_I^\dagger Y^J\Psi^{\dagger I}\Psi_{J}\nonumber\\&&-2\sum_{(I, J)=(1, 3),
(2, 4), (3, 2), (4, 1)}e^{-{i\over2}\pi
\gamma}Y_I^\dagger Y^J\Psi^{\dagger I}\Psi_{J}\nonumber\\&&+\sum_{(IJKL)=(1324),
(2413), (3241), (4132)}e^{{i\over2}\pi
i\gamma}Y_I^\dagger \Psi_{J}Y_{K}^\dagger \Psi_{L}\nonumber\\&&+\sum_{(IJKL)=(1423),
(2314), (3142), (4231)}e^{-{i\over2}\pi
i\gamma}Y_I^\dagger \Psi_{J}Y^{\dagger}_K\Psi_{L}\nonumber\\&&+\sum_{\text{other
terms}}\epsilon^{IJKL}Y_{I}^\dagger\Psi_J Y_K^+\Psi_{L}\Big]+
\text{h.c}\eea We list the non-vanishing phase factors multiplying
the second and the third terms of $V_F$ get changed in
tables~\ref{chargef1} and \ref{chargef2}.

\section{Two-loop anomalous dimensions in the scalar sector\label{twoloop}}
Now we compute the two-loop planar contributions to the anomalous
dimensions for the composite local operators in the scalar sector:
\begin{equation}\label{operator} {\cal O}^{I_1\cdots I_L}_{J_1\cdots J_L}\equiv\tr\left(Y^{I_1}Y^\dagger_{J_1}Y^{I_2}Y^\dagger_{J_2}\cdots Y^{I_L}Y^\dagger_{J_L}\right),\, L\geq 2. \end{equation}
The anomalous dimension matrix can be expressed as Hamiltonian
acting on an alternating $SU(4)$ spin chain with the spins at odd
lattice sides in the fundamental representation ($\bf{4}$) and the spins at even
lattices in the anti-fundamental representation ($\bf{\bar 4}$). The length of the spin
chain is $2L$. The involved Feynman diagrams are the same as the
ones in \cite{MinahanZarembo, Bak:2008cp}. 

\begin{figure}[h]
\centering
\includegraphics[width=4 cm]{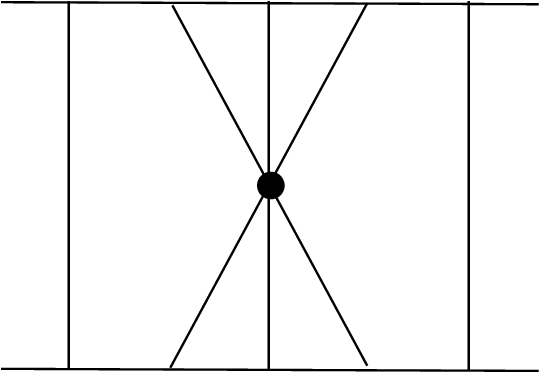}
\caption{The contribution to  anomalous dimension of ${\cal O}$  from two
loop contribution from scalar sextet interaction. In this context,
the horizontal lines represent the operators and the ordered
vertical lines denote the contraction between the two operators of
the fields included in trace.} \label{fig1}
\end{figure}

\begin{figure}[h]
\centering
\includegraphics[width=12 cm]{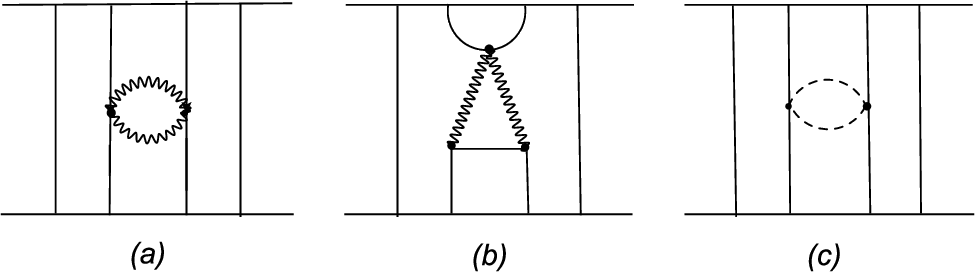}
\caption{The contribution to anomalous dimension of ${\cal O}$  from two
loop contribution of gauge and fermion exchange interaction. The
internal waved lines and dashed line stand gluon and scalar
respectively.} \label{fig2}
\end{figure}

Since only the third terms in $V_B$ is modified by the deformation,
one can see by some computations based on the Feynman diagram in
Fig.~\ref{fig1} that the Hamiltonian from $V_B$, \bea
H_B=\frac{\lambda^2}2\sum_{i=1}^{2L}[\mathbb{I}-2\mathbb{P}_{i,
i+2}-\mathbb{K}_{i, i+1}+\mathbb{P}_{i, i+2}\mathbb{K}_{i,
i+1}+\mathbb{K}_{i, i+1}\mathbb{P}_{i, i+1}]\,,\eea is now changed
into \bea
\widetilde{H}_B=\frac{\lambda^2}2\sum_{i=1}^{2L}[\mathbb{I}-2\widetilde{\mathbb{P}}_{i,
i+2}-\mathbb{K}_{i, i+1}+\mathbb{P}_{i, i+2}\mathbb{K}_{i,
i+1}+\mathbb{K}_{i, i+1}\mathbb{P}_{i, i+1}]\,,\eea where
$\lambda\equiv N/k$ is the 't Hooft coupling of ABJM theory, and
definition of $\mathbb{I}, \mathbb{P},  \mathbb{K}$ are
\begin{equation}\mathbb{I}^{IJ}_{KL}=\delta^I_K\delta^J_L,\, \mathbb{P}^{IJ}_{KL}=\delta^I_L\delta^J_K, \,
 \mathbb{K}^{IJ}_{KL}=\delta^{IJ}\delta_{KL}.\end{equation}
 The definition of $\widetilde{\mathbb{P}}_{i, i+2}$ is
\bea\left(\widetilde{\mathbb{P}}_{i, i+2}\right)^{I_i I_{i+1} I_{i+2}}_{J_i J_{i+1} J_{i+2}}&\equiv&\exp(-2\pi i\gamma(Q^{J_i}\times Q^{J_{i+1}}+Q^{J_{i+1}}
\times Q^{J_{i+2}}+Q^{J_{i+2}}\times Q^{J_i}))\nonumber\\
&\times&\left(\mathbb{P}_{i, i+2}\right)^{I_i I_{i+1} I_{i+2}}_{J_i
J_{i+1} J_{i+2}}.\label{new}\eea
\begin{figure}[h]
\centering
\includegraphics[width=12 cm]{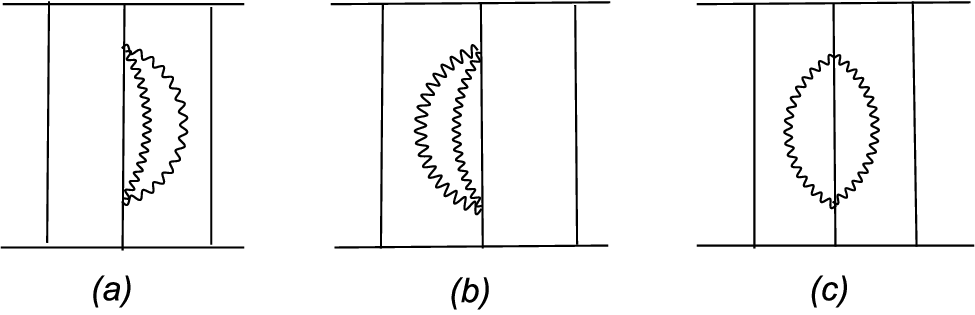}
\caption{The contribution to  wave function renormalization of $Y$,
$Y^+$  from two loop contribution of diamagnetic gauge interactions.
} \label{fig3}
\end{figure}

\begin{figure}[h]
\centering
\includegraphics[width=8 cm]{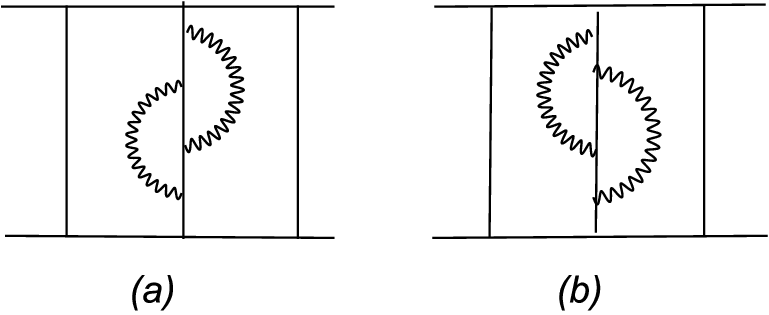}
\caption{The contribution to wave function renormalization of $Y$,
$Y^+$ is from two loop contribution of paramagnetic gauge
interactions.}\label{fig4}
\end{figure}

\begin{figure}[h]
\centering
\includegraphics[width=8 cm]{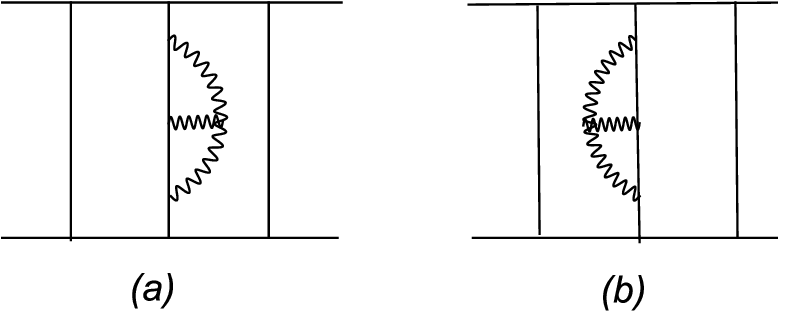}
\caption{The contribution to wave function renormalization of $Y$,
$Y^+$ is from two
loop contribution of Chern-Simons interaction.}\label{fig5} 
\end{figure}

\begin{figure}[h]
\centering
\includegraphics[width=12 cm]{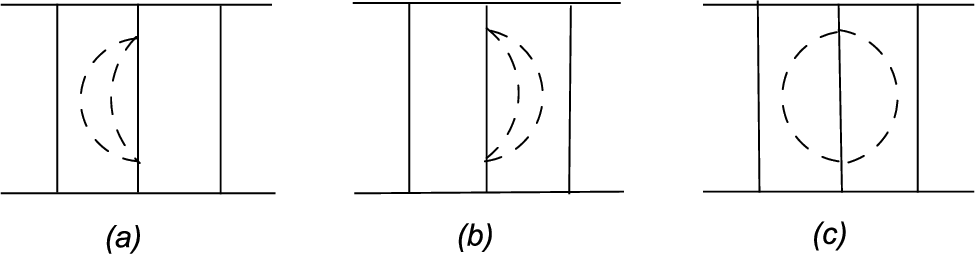}
\caption{The contribution to wave function renormalization of $Y$,
$Y^+$ is from two loop contribution of fermion pair interaction to
wave function
renormalization.}\label{fig6} 
\end{figure}

\begin{figure}[h]
\centering
\includegraphics[width=12 cm]{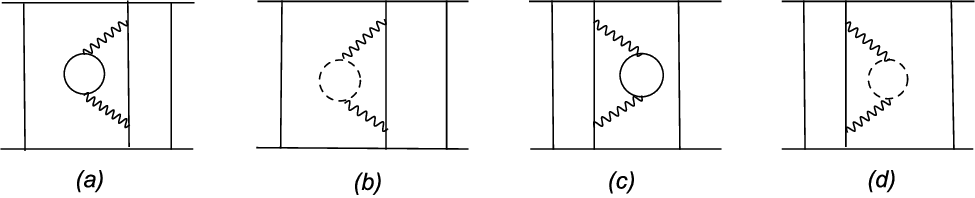}
\caption{The contribution to wave function renormalization of $Y$,
$Y^+$ is from two
loop contribution of vacuum polarization.}\label{fig7}
\end{figure}

 The contributions to the anomalous dimension of operator
(\ref{operator}) from gauge and fermion exchange interaction are
relevant to Feynman diagrams in Fig.~\ref{fig2}.  The wave function
renormalization of $Y, Y^+$ will also make contributions to
anomalous dimension of composite operators in eq.~(\ref{operator}).
There are three kinds of nonzero contribution arise from
interactions involving gauge boson loops, vertices given in
$V_F^{\text{deformed}}$ and gauge-matter interactions. The relevant
Feynman diagrams are given in Fig.~\ref{fig3}-Fig.~\ref{fig5},
Fig.~\ref{fig6} and Fig.~\ref{fig7}, respectively.
We find that these contributions \bea H_F&=&\lambda^2\sum_{i=1}^{2L}\mathbb{K}_{i, i+1},\\
H_{\rm gauge}&=&\lambda^2\sum_{i=1}^{2L}\left(-\frac14\mathbb{I}-\frac12\mathbb{K}_{i, i+1}\right),\\
H_Z&=&\lambda^2\sum_{i=1}^{2L}\frac34\mathbb{I},\eea are the same as the undeformed case. 

By summing over all of these contributions, we get\footnote{ Here
our convention is \bea
&&(\mathbb{P}_{i, i+2}\mathbb{K}_{i+1, i+2})^{K_iK_{i+1}K_{i+2}}_{J_iJ_{i+1}J_{i+2}}\nonumber\\
&=&(\mathbb{P}_{i, i+2})_{L_iL_{i+1}L_{i+2}}^{K_iK_{i+1}K_{i+2}}(\mathbb{K}_{i+1, i+2})^{L_iL_{i+1}L_{i+2}}_{J_iJ_{i+1}J_{i+2}}
\eea}
\bea \widetilde{H}_{\rm total}&=&\widetilde{H}_B+H_F+H_{gauge}+H_Z\nonumber\\
&=&\lambda^2\sum_{i=1}^{2L}\left(\mathbb{I}-\widetilde{\mathbb{P}}_{i,
i+2}+\frac12\mathbb{P}_{i, i+2}\mathbb{K}_{i, i+1}
+\frac12\mathbb{P}_{i, i+2}\mathbb{K}_{i+1,
i+2}\right).\label{hal}\eea

We notice that, as the undeformed case \cite{BakGangRey}, the
computations in the ABJ theory \cite{ABJ} with gauge group
$U(N)_k\times U(M)_{-k}$ is almost the same besides replacing the
factor $\lambda^2$ in the Hamiltonian by the factor
$\lambda\tilde\lambda$, where $\tilde\lambda$ is defined to be
$\tilde\lambda\equiv M/k$. So the computations and discussions in
the following applied to the scalar sector in ABJ theory at
two-loop level as well.

\section{The R matrices of deformed spin chain}
In this section, we will show that Hamiltonian obtained in the
previous section is integrable by constructing the R matrix which
satisfies the Yang-Baxter equation (YBE) and gives this Hamiltonian
through the transfer matrix by the standard procedure.

For the alternating spin chain, we need four R-matrices
$\widetilde{{\fR_{ij}^{\bf 4{4}}}}(u), \widetilde{{\fR_{ij}^{\bf
\bar{4}{4}}}}(u), \widetilde{{\fR_{ij}^{\bf 4{\bar{4}}}}}(u),
\widetilde{{\fR_{ij}^{\bf \bar{4}{\bar{4}}}}}(u)$ acting on the
space $V_i\otimes V_j$. Here the upper indices of
$\widetilde{{\fR_{ij}^{\bf 4{4}}}}(u)$ denote $SU(4)$ representations
related to the two spaces and $u$ denotes spectral parameter. We defined these R-matrices as
 \bea
{\widetilde{{\fR^{\bf 4{4}}}}}(u)^{IJ}_{KL}&=&\exp(i\pi\gamma(Q^J\times Q^I-Q^K\times Q^L)){{\fR^{\bf 4{4}}}}(u)^{IJ}_{KL},\label{tr1}\\
\widetilde{{\fR^{\bf 4\bar{4}}}}(u)^{IJ}_{KL}&=&\exp(-i\pi\gamma(Q^J\times Q^I-Q^K\times Q^L)){\fR^{\bf 4\bar{4}}}(u)^{IJ}_{KL},\\
\widetilde{{\fR^{\bf \bar{4}4}}}(u)^{IJ}_{KL}&=&\exp(-i\pi\gamma(Q^J\times Q^I-Q^K\times Q^L)){\fR^{\bf \bar{4}4}}(u)^{IJ}_{KL},\\
\widetilde{{\fR^{\bf \bar{4}\bar{4}}}}(u)^{IJ}_{KL}&=&\exp(i\pi\gamma(Q^J\times Q^I-Q^K\times Q^L)){\fR^{\bf
\bar{4}\bar{4}}}(u)^{IJ}_{KL},\label{tr4} \eea
where \bea {{\fR^{\bf 4{4}}}}(u)&=&u\mathbb{I}+\mathbb{P},\\
{{\fR^{\bf 4{\bar{4}}}}}(u)&=&-(u+2)\mathbb{I}+\mathbb{K},\\
{{\fR^{\bf \bar{4}{4}}}}(u)&=&-(u+2)\mathbb{I}+\mathbb{K},\\
{{\fR^{\bf \bar{4}{\bar{4}}}}}(u)&=&u\mathbb{I}+\mathbb{P}, \eea are
the R-matrices beform deformation \cite{MinahanZarembo, Bak:2008cp}.
From above formulas, we can get:
 \bea
{\widetilde{{\fR^{\bf 4{4}}}}}(u)^{IJ}_{KL}
&=&u\exp(2\pi i \gamma Q^J\times Q^I)\delta_K^I\delta_L^J+\delta_L^I\delta_K^J,\\
\widetilde{{\fR^{\bf 4\bar{4}}}}(u)^{IJ}_{KL}
&=&-(u+2)\exp(-2\pi i \gamma Q^J\times Q^I)\delta_K^I\delta_L^J+\delta^{IJ}\delta_{KL},\\
\widetilde{{\fR^{\bf \bar{4}4}}}(u)^{IJ}_{KL}&=&-(u+2)\exp(-2\pi i\gamma Q^J\times Q^I)\delta^I_K\delta^J_L+\delta^{IJ} \delta_{KL},\\
\widetilde{{\fR^{\bf
\bar{4}\bar{4}}}}(u)^{IJ}_{KL}&=&u\exp(2\pi i\gamma Q^J\times
Q^I)\delta_K^I\delta_L^J+\delta_L^I\delta_K^J.
 \eea
The R matrices before deformation satisfy YBE \cite{MinahanZarembo,
Bak:2008cp}: \bea{{\fR^{\bf 4{4}}}_{12}}(u-v){{\fR^{\bf
4{4}}}_{13}}(u){{\fR^{\bf 4{4}}}_{23}}(v)={{\fR^{\bf 4{4}}}_{23}}(v)
{{\fR^{\bf 4{4}}}_{13}}(u){{\fR^{\bf 4{4}}}_{12}}(u-v),\\
{{\fR^{\bf 4{4}}}_{12}}(u-v){{\fR^{\bf 4{\bar
4}}}_{13}}(u){{\fR^{\bf 4{\bar 4}}}_{23}}(v)={{\fR^{\bf 4{\bar
4}}}_{23}}(v)
{{\fR^{\bf 4{\bar 4}}}_{13}}(u){{\fR^{\bf 4{4}}}_{12}}(u-v),\\
{{\fR^{\bf \bar 4{\bar 4}}}_{12}}(u-v){{\fR^{\bf \bar
4{4}}}_{13}}(u){{\fR^{\bf \bar 4{4}}}_{23}}(v)={{\fR^{\bf\bar
4{4}}}_{23}}(v)
{{\fR^{\bf \bar 4{4}}}_{13}}(u){{\fR^{\bf \bar 4{\bar 4}}}_{12}}(u-v),\\
{{\fR^{\bf \bar 4{\bar 4}}}_{12}}(u-v){{\fR^{\bf \bar 4{\bar
4}}}_{13}}(u){{\fR^{\bf\bar 4{\bar4}}}_{23}}(v)={{\fR^{\bf
\bar4{\bar4}}}_{23}}(v) {{\fR^{\bf \bar
4{\bar4}}}_{13}}(u){{\fR^{\bf \bar4{\bar 4}}}_{12}}(u-v). \eea  As in \cite{Beisert:2005if}, the
choice of the phases in eqs.~(\ref{tr1}-\ref{tr4}) are such that the
R matrices after deformation still satisfy the YBE:
\bea{\widetilde{{\fR^{\bf 4{4}}}}}_{12}(u-v){\widetilde{{\fR^{\bf
4{4}}}}}_{13}(u){\widetilde{{\fR^{\bf
4{4}}}}}_{23}(v)={\widetilde{{\fR^{\bf 4{4}}}}}_{23}(v)
{\widetilde{{\fR^{\bf 4{4}}}}}_{13}(u){\widetilde{{\fR^{\bf 4{4}}}}}_{12}(u-v),\\
{\widetilde{{\fR^{\bf 4{4}}}}}_{12}(u-v){\widetilde{{\fR^{\bf 4{\bar 4}}}}}_{13}(u){\widetilde{{\fR^{\bf 4{\bar 4}}}}}_{23}(v)={\widetilde{{\fR^{\bf 4{\bar 4}}}}}_{23}(v){\widetilde{{\fR^{\bf 4{\bar 4}}}}}_{13}(u){\widetilde{{\fR^{\bf 4{4}}}}}_{12}(u-v),\\
{\widetilde{{\fR^{\bf \bar 4{\bar
4}}}}}_{12}(u-v){\widetilde{{\fR^{\bf \bar
4{4}}}}}_{13}(u){\widetilde{{\fR^{\bf \bar
4{4}}}}}_{23}(v)={\widetilde{{\fR^{\bf\bar 4{4}}}}}_{23}(v)
{\widetilde{{\fR^{\bf \bar 4{4}}}}}_{13}(u){\widetilde{{\fR^{\bf \bar 4{\bar 4}}}}}_{12}(u-v),\\
{\widetilde{{\fR^{\bf \bar 4{\bar
4}}}}}_{12}(u-v){\widetilde{{\fR^{\bf \bar 4{\bar
4}}}}}_{13}(u){\widetilde{{\fR^{\bf\bar
4{\bar4}}}}}_{23}(v)={\widetilde{{\fR^{\bf \bar4{\bar4}}}}}_{23}(v)
{\widetilde{{\fR^{\bf \bar 4{\bar4}}}}}_{13}(u){\widetilde{{\fR^{\bf
\bar4{\bar 4}}}}}_{12}(u-v). \eea

By introducing auxiliary spaces $V_0, V_{0^\prime}$, we can define the following
two transfer T-matrices\footnote{Comparing with the T-matrices in
\cite{Bak:2008cp}, we include the constant factor $2^{-L}$ as in
\cite{MinahanZarembo}.}:
 \bea \widetilde{T}_0 (u,a)=&&2^{-L}\widetilde{{\fR^{\bf 4{4}}}}_{01}(u) \widetilde{{\fR^{\bf 4\bar{4}}}}_{02}(u+a)
 \widetilde{{\fR^{\bf 4{4}}}}_{03}(u) \widetilde{{\fR^{\bf 4\bar{4}}}}_{04}(u+a)\cdots \nonumber\\
 &&\widetilde{{\fR^{\bf
4{4}}}}_{0(2L-1)}(u) \widetilde{{\fR^{\bf
4\bar{4}}}}_{0(2L)}(u+a)\,,\eea \bea\widetilde{ \overline{T}}_{0^{\prime}}
(u,\bar{a}) =&& 2^{-L} \widetilde{{\fR^{\bf\bar{
4}{4}}}}_{0^\prime1}(u+\bar{a}) \widetilde{{\fR^{\bf\bar{
4}\bar{4}}}}_{0^\prime2}(u) \widetilde{{\fR^{\bf\bar{
4}{4}}}}_{0^\prime3}(u+\bar{a}) \widetilde{{\fR^{\bf\bar{
4}\bar{4}}}}_{0^\prime4}(u)\cdots \nonumber\\&& \widetilde{{\fR^{\bf\bar{
4}{4}}}}_{0^\prime(2L-1)}(u+\bar{a}) \widetilde{{\fR^{\bf\bar{
4}\bar{4}}}}_{0^\prime(2L)}(u) \,. \eea

The YBE will lead to the following relations:
\bea\widetilde{{\fR^{\bf 4{4}}}}_{00^\prime}(\nu-\mu)
\widetilde{T}_0(\nu, a) \widetilde{T}_{0^\prime}(\mu,
a)&=&\widetilde{T}_{0^\prime}(\mu, a)\widetilde{T}_0(\nu,
a)\widetilde{{\fR^{\bf
4{4}}}}_{00^\prime}(\nu-\mu),\\
\widetilde{{\fR^{\bf \bar{4}{\bar{4}}}}}_{00^\prime}(\nu-\mu)
\widetilde{\overline{T}}_0(\nu, \bar{a})
\widetilde{\overline{T}}_{0^\prime}(\mu,
\bar{a})&=&\widetilde{\overline{T}}_{0^\prime}(\mu,
\bar{a})\widetilde{\overline{T}}_0(\nu, \bar{a})\widetilde{{\fR^{\bf
\bar{4}{\bar{4}}}}}_{00^\prime}(\nu-\mu),\\
 \widetilde{{\fR^{\bf
4{\bar{4}}}}}_{00^\prime}(\nu-\mu+a) \widetilde{T}_0(\nu, a)
\widetilde{\overline{T}}_{0^\prime}(\mu,
-a)&=&\widetilde{\overline{T}}_{0^\prime}(\mu, -a)\widetilde{T}_0(\nu,
a)\widetilde{{\fR^{\bf 4{\bar{4}}}}}_{00^\prime}(\nu-\mu+a).\eea

Then the traces of the two T-matrices
\bea \widetilde{\tau}(\nu, a)&=&\tr_0 {\widetilde{T}}_0 (\nu, {a})\,,\\
\widetilde{\overline{\tau}}(\nu, a)&=&\tr_{0^\prime}
{\widetilde{\overline{T}}}_{0^\prime}(\nu, {a})\,, \eea satisfy \bea &&
[\widetilde{\tau}(\nu, a), \widetilde{\tau}(\mu, a)]=0
\nonumber\\
&& [\widetilde{\overline\tau}(\nu,
\bar{a}),\widetilde{\overline\tau}(\mu, \bar{a})]=0\,, \nonumber\\&&
[\widetilde{\tau}(\nu, a), \widetilde{\overline\tau}(\mu, -{a})]=0
\,.
\eea From now on, we restrict $a$ to be purely imaginary so that
$\widetilde{\tau}(\nu, a)$ and $\widetilde{\overline\tau}(\mu,
\bar{a})$ commute with each other.

Now we compute the Hamiltonian from the R-matrices: \bea H
=\frac{\partial \log \widetilde{\tau}(u, a)}{\partial
u}\big{|}_{u=0}+\frac{\partial \log \widetilde{\bar\tau}(u,
\bar{a})}{\partial u}\big{|}_{u=0}. \label{Hbeta}\eea

After some computations, we get \bea H=\sum_{i=1}^{2L}H_i, \eea with
\bea H_{2l-1}&=&\frac{1}{a^2-4}((a-2)\mathbb{I}+(a^2-4)\widetilde{\mathbb{P}}_{2l-1, 2l+1}\nonumber\\
&-&(a-2)\mathbb{P}_{2l-1, 2l+1}\mathbb{K}_{2l-1, 2l}+(a+2)\mathbb{P}_{2l-1, 2l+1}\mathbb{K}_{2l, 2l+1})\,,\\
H_{2l}&=&\frac{1}{a^2-4}(-(a+2)\mathbb{I}+(a^2-4)\widetilde{\mathbb{P}}_{2l, 2l+2}\nonumber\\
&+&(a+2)\mathbb{P}_{2l, 2l+2}\mathbb{K}_{2l,
2l+1}-(a-2)\mathbb{P}_{2l, 2l+2}\mathbb{K}_{2l+1, 2l+2})\,. \eea Here we have already used the relation $\bar{a}=-a$.
 The details of the computations are deferred to the Appendix. After setting $a=0$, multiplying $H_i$ by $-1$, then  shifting $H_i$ by $\frac32\mathbb{I}$,
the above Hamiltonian coincides with the one in the previous section
obtained from the perturbative computations in field theory side.

\section{Eigenvalues of deformed spin chain Hamiltonian and Bethe ansatz equations}
In previous section, we constructed transfer matrix and Hamiltonian
of the deformed spin chain. We now derive the Bethe ansatz equation
through diagonalizing the transfer matrices. By choosing the ground
state or highest-weight state as $|1\bar{4}1\bar{4}\cdots>$ and
introducing three sets of Bethe roots $(l_a, m_b, r_c), 1\le a\le
N_l, 1\le b\le N_m, 1\le c\le N_r$, we can get the eigenvalues of
$\tilde\tau(\nu, 0)$ as
\bea\widetilde{\Lambda}(\nu)&=&2^{-L}(\nu+1)^L(-\nu-2)^L\exp(-{i\over
2}\pi \gamma L-{i\over 2}\gamma N_m+i\pi \gamma
N_r)\prod_{a=1}^{N_l}{\nu+il_a-{1\over 2}\over \nu+il_a+{1\over
2}}\nonumber\\&&+2^{-L}(\nu+1)^L(-\nu)^L\exp(-{i\over 2}\pi \gamma
L-{i\over 2}\gamma N_m+i\pi \gamma
N_l)\prod_{c=1}^{N_r}{\nu+ir_c+{5\over 2}\over \nu+ir_c+{3\over
2}}\nonumber\\&&+2^{-L}(-\nu-2)^L\nu^L\exp({i\over 2}\pi\gamma
L+{i\over 2}\pi \gamma N_m-i\pi \gamma
N_r)\prod_{a=1}^{N_l}{\nu+il_a+{3\over 2}\over \nu+il_a+{1\over
2}}\prod_{b=1}^{N_m}{\nu+im_b\over
\nu+im_b+1}\nonumber\\&&+2^{-L}(-\nu-2)^L\nu^L\exp({i\over
2}\pi\gamma L+{i\over 2}\pi \gamma N_m-i\pi\gamma
N_l)\prod_{c=1}^{N_c}{\nu+ir_c+{1\over 2}\over \nu+ir_c+{3\over
2}}\prod_{b=1}^{N_m}{\nu+im_b+2\over \nu+im_b+1}.\nonumber\\\eea

As the undeformed case \cite{MinahanZarembo, Bak:2008cp}, $N_l, N_m, N_r$ should satisfy:
\begin{equation} 2N_l\le L+N_m,\, 2N_r\le L+N_m,\, 2N_m\le N_l+N_r.\end{equation}

Similarly, we can get the eigenvalues of $\tilde{\bar{\tau}}(\nu,
0)$ 
as \bea
\widetilde{\bar{\Lambda}}(\nu)&=&2^{-L}(-\nu)^L(\nu+1)^L\exp({i\over
2}\pi \gamma L+{i\over 2}\pi \gamma N_m-i\pi \gamma
N_r)\prod_{a=1}^{N_l}{\nu+il_a+{5\over 2}\over \nu+il_a+{3\over
2}}\nonumber\\&&+2^{-L}(\nu+1)^L(-\nu-2)^L\exp({i\over 2}\pi\gamma
L+{i\over 2}\pi \gamma N_m-i\pi \gamma
N_l)\prod_{c=1}^{N_r}{\nu+ir_c-{1\over 2}\over \nu+ir_c+{1\over
2}}\nonumber\\&& +2^{-L}(-\nu-2)^L\nu^L\exp(-{i\over 2}\pi \gamma
L-{i\over 2}\pi \gamma N_m+i\pi\gamma
N_r)\prod_{a=1}^{N_l}{\nu+il_a+{1\over 2}\over \nu+il_a+{3\over
2}}\prod_{b=1}^{N_m}{\nu+im_b+2\over
\nu+im_b+1}\nonumber\\&&+2^{-L}(-\nu-2)^L\nu^L\exp(-{i\over 2}\pi
\gamma L-{i\over 2}\pi \gamma N_m+i\pi \gamma
N_l)\prod_{b=1}^{N_m}{\nu+im_b\over\nu+im_b+1}\prod_{c=1}^{N_r}{\nu+ir_c+{3\over
2}\over \nu+ir_c+{1\over 2}}.\nonumber\\\eea

By demanding the residue vanishes at every pole of
$\widetilde{\Lambda}(\nu)$, we get the following set of Bethe Ansatz
equations.

\bea \exp(i\pi\gamma L+i\pi \gamma
N_m-2\pi i \gamma N_r)\left({l_a-{i\over 2}\over l_a+{i\over
2}}\right)^L=\prod_{a'\neq a}{l_a-l_{a'}-i\over
l_a-l_{a'}+i}\prod_{b=1}^{N_m}{l_a-m_b+{i\over 2}\over
l_a-m_b-{i\over 2}},\nonumber\\
\exp(-i\pi \gamma L-i\pi\gamma N_m+2\pi i \gamma
N_l)\left({r_c-{i\over 2}\over r_c+{i\over
2}}\right)^L=\prod_{b=1}^{N_m}{r_c-m_b+{i\over 2}\over
r_c-m_b-{i\over 2}}\prod_{c'\neq c}{r_c-r_{c'}-i\over
r_c-r_{c'}+i},\nonumber\\
\exp(-i\pi \gamma N_l+i\pi \gamma
N_r)=\prod_{a=1}^{N_l}{m_b-l_a+{i\over 2}\over m_b-l_a-{i\over
2}}\prod_{b\neq b'}{m_b-m_{b'}-i\over m_b-m_{b'}+i} \prod_{c=1}^{N_r}{m_b-r_c+{i\over 2}\over m_b-r_c-{i\over
2}}.
\label{Betheansatz}\eea   We will get the same set of equations if we
start with $\widetilde{\bar{\Lambda}}(\nu)$ instead. It is a
consistent check that the same set of Bethe ansatz equations
remove potential simple pole terms for $\widetilde{\Lambda}(\nu)$
and $\widetilde{\bar{\Lambda}}(\nu)$. One can see the equations just above go back to Bethe ansatz equations given in
\cite{MinahanZarembo, Bak:2008cp} by setting $\gamma=0$.

From the above eigenvalues, we can get the total momentum as\footnote{Here the fundamental domain of the momentum is chosen to
be $[0, 2\pi)$.}\bea
P_{total}&=&{1\over
i}\left[\log\widetilde{\Lambda}(0)+\log\widetilde{\bar{\Lambda}}(0)\right]\nonumber\\
&=&{1\over i} \left[i\pi \gamma N_r-i\pi \gamma
N_l+\sum_{a=1}^{N_l}\log{il_a-{1\over 2}\over il_a+{1\over
2}}+\sum_{c=1}^{N_r}\log{ir_c-{1\over 2}\over ir_c+{1\over
2}}\right]. \eea

 Notice that
 \bea (\tau(0, 0)
\bar{\tau(0, 0)})_{I_1\cdots I_{2L}}^{J_1\cdots J_{2L}}=\delta_{I_3}^{J_1}\delta_{I_4}^{J_2}\cdots \delta_{I_1}^{J_{2L-1}}\delta_{I_2}^{J_{2L}}. \eea
 acts trivially on the physics state.
 The total momentum should vanish,  then we have the following constraint,
 \bea 1=\exp(i\pi \gamma
N_r-i\pi \gamma N_l)\prod_{a=1}^{N_l}{il_a-{1\over 2}\over
il_a+{1\over 2}}\prod_{c=1}^{N_r}{ir_c-{1\over 2}\over ir_c+{1\over
2}}.\eea

The total energy (the eigenvalue of ${\tilde H}_{total}$ in eq.~(\ref{hal})) is  \bea E_{total}&=&3L-\left[{d\over d\nu }\log
\widetilde{\Lambda}(\nu)+{d\over d\nu}\log
\widetilde{\bar{\Lambda}}(\nu)\right]\Big|_{\nu=0}\nonumber\\
&=&\sum_{a=1}^{N_l}{1\over l_a^2+{1\over
4}}+\sum_{c=1}^{N_r}{1\over r_c^2+{1\over 4}}.\eea

From the total energy, we can get the eigenvalues of the anomalous dimension matrix of the operators in the scalar
sector eq.~(\ref{operator})
in the $\beta$-deformed ABJM theory.

\section{The non-supersymmetric three-parameter $\gamma$-deformation of ABJ(M) theory}
The three-parameter deformation can be
performed by replacing all of the ordinary product $fg$ of two
fields $f$ and $g$ in the Lagrangian by the following star product \cite{Imeroni:2008cr}:
\begin{equation}
     f * g = e^{i \pi \gamma_{i} Q^f_j Q^g_k\epsilon^{ijk} } f g\,,
\end{equation}
where $Q^f_i, i=1, 2,3$ are three global $U(1)$ charges carrying by
the field $f$ and $\gamma_i$'s are three real deformation parameters. We choose the
$U(1)_i$ charges for the scalars and the fermions as in
table~\ref{threeparamtercharges}. The gauge field is neutral under
these symmetries. One can see that this deformation degenerates
to the one of $\beta$-deformation by setting deformation parameters
$\gamma_1=\gamma_2=0, \gamma_3=\gamma$.

After analogous calculation as in section~\ref{twoloop}, one can
find that in the  scalar sector and at the two-loop level, the
Hamiltonian is still given by eq.~(\ref{hal}) with $\gamma$ in eq.~(\ref{new}) being $\gamma_3$. The differences between
$\beta$- and $\gamma$-deformations may appear at higher loop
orders or in other sectors of the theories.

\begin{table}
\centering \caption{{$U(1)^3$ charges of the scalars and fermions of
the ABJM theory used for $\gamma$-deformation.}}\label{threeparamtercharges}
\begin{tabular}{|c|c|c|c|c|c|c|c|c|}
\hline
& $Y^1$ & $Y^2$ & $Y^3$ & $Y^4$ & $\Psi^{\dagger 1}$ & $\Psi^{\dagger 2}$ & $\Psi^{\dagger 3}$ & $\Psi^{\dagger 4}$\\
\hline
$U(1)_1$ & $\frac{1}{2}$ & $-\frac{1}{2}$ & $0$ & $0$ & $\frac{1}{2}$& $-\frac{1}{2}$& $0$& $0$ \\
\hline
$U(1)_2$ & $0$ & $0$ & $\frac{1}{2}$ & $-\frac{1}{2}$ & $0$ & $0$ & $\frac{1}{2}$ & $-\frac{1}{2}$ \\
\hline
$U(1)_3$ & $\frac{1}{2}$ & $\frac{1}{2}$ & $\frac{1}{2}$ & $\frac{1}{2}$ & $-\frac{1}{2}$ & $-\frac{1}{2}$ & $-\frac{1}{2}$ & $-\frac{1}{2}$ \\
\hline
\end{tabular}
\end{table}


\section{Conclusion and Discussions}

In this note, we start a study on the integrable structure of
$\beta$- and $\gamma$-deformed ABJ(M) theory, beginning with the
scalar sector at the two-loop level in the planar limit. We first
perform perturbative computations of the anomalous dimension matrix
and express the result as a Hamiltonian acting on alternating
$SU(4)$ spin chain. We find that only one term in the Hamiltonian is
deformed and that the differences between $\beta$-deformation and
$\gamma$-deformation are invisible in this sector at two loop level.
As the undeformed case, the difference between  Hamiltonian for
deformed  ABJM theory and deformed ABJ theory only appears in the
prefactor. So in this sector and at this order of the perturbation
theory, the violation of parity invariance in the deformed ABJ
theory does not affect the integrability. Based on the structure of
the deformations, we choose a suitable deformation  of the
R-matrices. Then the Bethe ansatz equations are obtained through
diagonalizing the transfer matrices.

There are several directions worth pursuing. The study
here in the field theory side can be extended to full sector and/or
higher loop order as in the undeformed case
\cite{Zwiebel:2009vb}-\cite{Bak:2009tq}. It is also interesting to
reproduce the Bethe ansatz equation starting from  the S-matrix of
the spin chain based on the studies in \cite{Ahn:2008aa, Arutyunov:2010gu, Ahn:2010ws,
Ahn:2012hs}. In the string theory side, one could try to construct
the Lax pair and the infinite number of conserved currents on the worldsheet.
Even for
the undeformed case, the story of IIA string on $AdS_4\times CP^3$
has already been much richer than the one of IIB string on
$AdS_5\times S^5$, partly because that now the $OSp(6|4)/U(3)\times
SO(1, 3)$ coset action can only describe a subset of the complete
Green-Schwarz action \cite{arXiv:0811.1566, arXiv:1009.3498,
arXiv:1101.3777}.

\section*{Acknowledgements}
The authors are grateful to Dongsu Bak, Bin Chen, Peng Gao, Miao Li, Xiao Liu,
Jian-Xin Lu, Jian-Feng Wu, Zhi-Guang Xiao, Jie Yang and Hossein Yavartanoo
for useful discussions. JW would like to thank SISSA, ICTS-USTC and
School of Mathematical Sciences, USTC for warm hospitality. He is
also very thankful to the organizers and participants of Symposium
on Strings and Particle Physics 2012 held at CTP, SCU and International
Workshop on Quantum Aspects of Black Holes held in CQUeST, Sogang
University. This work was supported in part by the National Natural
Science Foundation of China (No.10821504 (SH), No.10975168 (SH),
No.11035008 (SH), No. 11105154 (JW), and No. 11222549 (JW)), and in
part by the Ministry of Science and Technology of China under Grant
No. 2010CB833004. SH also would like to appreciate the general
financial support from China Postdoctoral Science Foundation No.
2012M510562. JW gratefully acknowledges the support of K.~C.~Wong
Education Foundation and Youth Innovation Promotion Association, CAS
as well.

\appendix
\section{Hamiltonian from deformed R-matrices}

In this appendix, we compute the Hamiltonian of deformed spin chain
in eq.~(\ref{Hbeta}).
 For this, we should
compute $\widetilde{\tau}'(0, a)$, where the prime denotes the
derivative with respect to spectrum parameter $u$. \bea
\widetilde{\tau}'(0, a)&= &2^{-L}\tr \sum_{i} \widetilde{\fR^{\bf
44}}_{01}(0)...{d \widetilde{\fR^{\bf 44}}_{0(2i-1)}(u)\over d
u}|_{u=0}...\widetilde{\fR^{\bf 4\bar{\bf
4}}}_{0(2L)}(a)\nonumber\\
&+&2^{-L}\tr \sum_{i}\widetilde{\fR^{\bf
44}}_{01}(0)...{d\widetilde{\fR^{\bf 4\bar{\bf
4}}}_{0(2i)}(u+a)\over d u}|_{u=0}...\widetilde{\fR^{\bf 4\bar{\bf
4}}}_{0(2L)}(a).\label{deformedtauprime}\eea Where the spectrum
parameter $u$ has been set as vanishing.
 The $i$-th term in  first part of eq.~(\ref{deformedtauprime}) can be written down as following \bea &&
2^{-L}\left(\mathbb{P}_{01}\right)_{K_1 J_1}^{K_0
I_1}\widetilde{\fR^{\bf 4\bar{4}}}_{02}(a)_{K_2 J_2}^{K_1
I_2}...\widetilde{\fR^{\bf 4\bar{4}}}_{0 (2i-2)}(a)_{K_{2i-2}
J_{2i-2}}^{K_{2i-3}I_{2i-2}}\delta_{K_{2i-1}}^{K_{2i-2}}e^{-2\pi i\gamma Q^{K_{2i-1}}\times Q^{I_{2i-1}}}\nonumber\\
&&\delta_{J_{2i-1}}^{I_{2i-1}}\widetilde{\fR^{\bf 4\bar{4}}}_{0
(2i)}(a)_{K_{2i} J_{2i}}^{K_{2i-1}I_{2i}}...
\left(\mathbb{P}_{0(2L-1)}\right)_{K_{2L-1} J_{2L-1}}^{K_{2L-2}
I_{2L-1}}{\widetilde{\fR^{\bf 4\bar{4}}}}_{0(2L)}(a)^{K_{2L-1}I_{2L}}_{K_{0}J_{2L}}\nonumber\\
&=&2^{-L} \widetilde{\fR^{\bf 4\bar{4}}}_{02}(a)_{J_3 J_2}^{I_1
I_2}...\widetilde{\fR^{\bf 4\bar{4}}}_{0 (2i-2)}(a)_{K_{2i-1}
J_{2i-2}}^{I_{2i-3}I_{2i-2}}e^{-2\pi i\gamma Q^{K_{2i-1}}\times
Q^{I_{2i-1}}}\delta_{J_{2i-1}}^{I_{2i-1}}\widetilde{\fR^{\bf
4\bar{4}}}_{0 (2i)}(a)_{J_{2i+1}
J_{2i}}^{K_{2i-1}I_{2i}}\nonumber\\&&...{\widetilde{\fR^{\bf
4\bar{4}}}_{0(2L)}}(a)^{I_{2L-1}I_{2L}}_{J_{1}J_{2L}}.\eea

The $i$-th term in the second part of eq.~(\ref{deformedtauprime}) is \bea &&
2^{-L}\left(\mathbb{P}_{01}\right)_{K_1 J_1}^{K_0
I_1}\widetilde{\fR^{\bf 4\bar{4}}}_{02}(a)_{K_2 J_2}^{K_1
I_2}\cdots\widetilde{\fR^{\bf 4\bar{4}}}_{0 (2i-2)}(a)_{K_{2i-2}
J_{2i-2}}^{K_{2i-3}I_{2i-2}}\left(\mathbb{P}_{0(2i-1)}\right)_{K_{2i-1}J_{2i-1}}^{K_{2i-2}I_{2i-1}}\left(-\mathbb{I}\right)_{K_{2i}J_{2i}}^{K_{2i-1}I_{2i}}
\nonumber\\& &e^{2\pi i\gamma Q^{K_{2i}}\times Q^{I_{2i}}}
\left(\mathbb{P}_{0(2i+1)}\right)_{K_{2i+1}J_{2i+1}}^{K_{2i}I_{2i+1}}
\widetilde{\fR^{\bf 4\bar{4}}}_{0 (2i+2)}(a)_{K_{2i+2}
J_{2i+2}}^{K_{2i+1}I_{2i+2}}...\widetilde{\fR^{\bf 4\bar{4}}}_{0
2L}(a)_{K_0J_{2L}}^{K_{2L-1}I_{2L}}\nonumber\\
&=& -2^{-L}\widetilde{\fR^{\bf 4\bar{4}}}_{02}(a)_{J_3 J_2}^{I_1
I_2}\cdots\widetilde{\fR^{\bf 4\bar{4}}}_{0(2i-2)}(a)_{J_{2i-1}
J_{2i-2}}^{I_{2i-3}
I_{2i-2}}\delta_{J_{2i}}^{I_{2i}}\delta_{J_{2i+1}}^{I_{2i-1}}e^{2\pi i\gamma
Q^{I_{2i-1}}\times Q^{I_{2i}}}
\nonumber\\&&\widetilde{\fR^{\bf
4\bar{4}}}_{0(2i+2)}(a)_{J_{2i+3} J_{2i+2}}^{I_{2i+1}
I_{2i+2}}\cdots\widetilde{\fR^{\bf 4\bar{4}}}_{0 (2L)}(a)_{J_1
J_{2L}}^{I_{2L-1} I_{2L}}.\eea

The deformed $\widetilde{\tau}^{-1}(0, a)$ is \bea\widetilde{\tau}^{-1}(0, a)&=&
2^L\left[{\widetilde{\fR^{\bf
4\bar{4}}}_{0(2L)}}^{-1}(a)\right]^{J_1J_{2L}}_{I_{2L-1}I_{2L}}...\left[{\widetilde{\fR^{\bf
4\bar{4}}}_{0(2)}}^{-1}(a)\right]^{J_{3}J_{2}}_{I_{1}I_{2}},\eea
where \bea \left[{\widetilde{\fR^{\bf
4\bar{4}}}_{0(2i)}}^{-1}(a)\right]^{J_{2i-1}J_{2i}}_{I_{2i+1}I_{2i}}=-\left({1\over
a+2} \mathbb{I}^{J_{2i-1} J_{2i}}_{I_{2i+1}I_{2i}} e^{-2\pi i\gamma
Q^{J_{2i-1}}\times Q^{J_{2i}}}+{1\over a^2-4}\mathbb{K}^{J_{2i-1}
J_{2i}}_{I_{2i+1}I_{2i}}\right).\eea One can use above formula to
obtain that \bea
&&\widetilde{\tau}^{-1}(u, a)\widetilde{\tau}'(u, a)|_{u=0}\nonumber\\
&=&\sum_{i=1}^L \mathbb{I}^{K_1\cdots K_{2i-3}}_{J_1\cdots
J_{2i-3}}\otimes\left[{\widetilde{\fR^{\bf
4\bar{4}}}_{0(2i)}}^{-1}(a)\right]^{K_{2i+1}K_{2i}}_{I_{2i-1}I_{2i}}\left[{\widetilde{\fR^{\bf
4\bar{4}}}_{0(2i-2)}}^{-1}(a)\right]^{K_{2i-1}K_{2i-2}}_{I_{2i-3}I_{2i-2}}\nonumber\\&&e^{-2\pi i\gamma
Q^{\widetilde{K}_{2i-1}}\times
Q^{I_{2i-1}}}\delta_{J_{2i-1}}^{I_{2i-1}}{\widetilde{\fR^{\bf
4\bar{4}}}_{0(2i-2)}}(a)^{I_{2i-3}I_{2i-2}}_{\tilde{K}_{2i-1}J_{2i-2}}{\widetilde{\fR^{\bf
4\bar{4}}}_{0(2i)}}(a)^{\tilde{K}_{2i-1}I_{2i}}_{J_{2i+1}J_{2i}}
\otimes\mathbb{I}^{K_{2i+2}\cdots K_{2L}}_{J_{2i+2}\cdots
J_{2L}}\nonumber\\&+&\sum_{i=1}^L\mathbb{I}^{K_1\cdots
K_{2i-1}}_{J_1\cdots J_{2i-1}}\otimes \left[{\widetilde{\fR^{\bf
4\bar{4}}}_{0(2i)}}^{-1}(a)\right]^{K_{2i+1}K_{2i}}_{I_{2i-1}I_{2i}}\left(-\delta^{I_{2i-1}}_{J_{2i+1}}\delta^{I_{2i}}_{J_{2i}}\right)e^{2\pi i\gamma
Q^{I_{2i-1}}\times Q^{I_{2i}}}\otimes\mathbb{I}^{K_{2i+2}\cdots
K_{2L}}_{J_{2i+2}\cdots J_{2L}}\nonumber\\&=&\sum_{i=1}^L
\mathbb{I}^{K_1\cdots K_{2i-3}}_{J_1\cdots
J_{2i-3}}\otimes\left[{\widetilde{\fR^{\bf
4\bar{4}}}_{0(2i)}}^{-1}(a)\right]^{K_{2i+1}K_{2i}}_{J_{2i-1}I_{2i}}\left[{\widetilde{\fR^{\bf
4\bar{4}}}_{0(2i-2)}}^{-1}(a)\right]^{K_{2i-1}K_{2i-2}}_{I_{2i-3}I_{2i-2}}e^{-2\pi i\gamma
Q^{\widetilde{K}_{2i-1}}\times
Q^{J_{2i-1}}}\nonumber\\&&{\widetilde{\fR^{\bf
4\bar{4}}}_{0(2i-2)}}(a)^{I_{2i-3}I_{2i-2}}_{\tilde{K}_{2i-1}J_{2i-2}}{\widetilde{\fR^{\bf
4\bar{4}}}_{0(2i)}}(a)^{\tilde{K}_{2i-1}I_{2i}}_{J_{2i+1}J_{2i}}
\otimes\mathbb{I}^{K_{2i+2}\cdots K_{2L}}_{J_{2i+2}\cdots
J_{2L}}\nonumber\\&-&\sum_{i=1}^L\mathbb{I}^{K_1\cdots
K_{2i-1}}_{J_1\cdots J_{2i-1}}\otimes \left[{\widetilde{\fR^{\bf
4\bar{4}}}_{0(2i)}}^{-1}(a)\right]^{K_{2i+1}K_{2i}}_{J_{2i+1}J_{2i}}e^{2\pi i\gamma
Q^{J_{2i+1}}\times Q^{J_{2i}}}\otimes\mathbb{I}^{K_{2i+2}\cdots
K_{2L}}_{J_{2i+2}\cdots J_{2L}}\nonumber\\&=&
\sum_{i=1}^L\frac{1}{a^2-4}((a-2)\mathbb{I}+(a^2-4)\widetilde{\mathbb{P}}_{2i-1,
2i+1}-(a-2)\mathbb{P}_{2i-1, 2i+1}\mathbb{K}_{2i-1, 2i}
\nonumber\\
&+&(a+2)\mathbb{P}_{2i-1, 2i+1}\mathbb{K}_{2i, 2i+1})
\label{detail-Ham}\eea Similarly, one can get: \bea
&&\widetilde{\overline{\tau}}^{-1}(u, a)\widetilde{\overline{\tau}}'(u, a)|_{u=0}\nonumber\\
&=&\sum_{i=1}^{L}\frac{1}{a^2-4}(-(a+2)\mathbb{I}+(a^2-4)\widetilde{\mathbb{P}}_{2i, 2i+2}+(a+2)\mathbb{P}_{2i, 2i+2}\mathbb{K}_{2i, 2i+1}\nonumber\\
&-&(a-2)\mathbb{P}_{2i, 2i+2}\mathbb{K}_{2i+1, 2i+2}), \eea where we have use the fact that $a$ is purely imaginary. From
these two equations, we can get the Hamiltonian given in the main
text.

\end{document}